\documentclass[11pt,a4paper]{article}

\usepackage[T1]{fontenc}
\usepackage[utf8]{inputenc}
\usepackage[margin=1in]{geometry}
\usepackage{booktabs}
\usepackage{tikz}
\usepackage{hyperref}
\hypersetup{hidelinks}

\newcommand{\companionid}{}
\newcommand{\companionnote}{\if\relax\detokenize\expandafter{\companionid}\relax\else\space(arXiv:\companionid)\fi}

\title{%
  Empty Intersection:\\
  \large Provenance Coverage Rose to 98\% and Neither Verification
  Decision Moved%
}

\author{Dong Hyeon Jeon\\\small Independent Researcher}
\date{}

\begin{document}
\maketitle

%
\begin{abstract}
Two structural defenses for provenance---a grade on every row, so that a
verification routine cannot mistake the system's own output for an observation,
and a single write ingress, so that the grade is enforced rather than merely
conventional---were measured against the production deployment that motivated
them, over a frozen snapshot of 194{,}620 rows and the two verification
decisions the snapshot supports. Neither reaches either decision. Both were
prescribed by a companion paper\companionnote{}, which diagnosed that
deployment: its verification routines decided outcomes using values the system
itself had written. Neither prescription is new:
both are established practice in fields that do not cite one another, and no
prior work measuring whether either changes a verdict was found, so what is
offered here is the measurement and not the prescriptions. Filtering the
verification queries by grade turns both decisions from pass to undetermined;
widening the grade vocabulary raises classified coverage from $36.1\%$ to
$98.4\%$; a single ingress requiring a grade refuses $3{,}070$ writes. None
of the three gives either decision admissible input. The prescriptions do not
fail at what they specify. Each is stated over the population and makes no
reference to any decision, so neither says which rows a decision will read, and
the rows each intervention repairs and the 32 rows the decisions read do not
intersect. The intervention that changed the most rows shows the reach most
plainly: all $121{,}296$ rows it moved from unnameable to named fall outside
both query windows. This paper reports the conditions, measured rather than
designed, under which the decisions would have admissible input at all, and
notes that the two decisions are blocked for different reasons.
\end{abstract}

\section{Introduction}
\label{sec:introduction}

A companion paper\companionnote{} documented nine cases in which a description
of a system was taken as evidence for the system's state, and it ended with two
prescriptions. Mark every row with a provenance grade, so that a verification
routine cannot mistake the system's own output for an observation. Route every
write through a single ingress, so that the grade is enforced rather than merely
conventional.

This paper measures those prescriptions against the deployment that produced them.

The setting is favourable to the prescriptions in every respect that matters. The
diagnosis and the prescriptions come from the same body of work, so nothing is lost
in translation. The measurements run against a frozen snapshot of that deployment,
so the prescriptions are tested on the exact state that motivated them rather than
on a reconstruction. And the defect they target is present and severe: of the
values one verification decision rests on, \textbf{24 of 24} were written by the
system itself, and its relational counterpart reads \textbf{0 of the 45} rows it
had just produced while reporting success.

Three interventions were measured. Filtering the verification queries so they
consume only observation-graded rows turns both decisions from pass to
undetermined. Widening the grade vocabulary raises classified coverage from
$36.1\%$ to $98.4\%$ of the population (Table~\ref{tab:i1} states both to four
decimal places). A single write ingress that requires a grade refuses
$3{,}070$ of $194{,}620$ recorded writes.

\textbf{None of the three gives either decision admissible input.}

That is the result, and it is worth stating precisely, because it is easy to
overread in both directions. It is not a finding that provenance grades or write
gateways are without value; nothing here measures what the refused writes or the
newly nameable rows are worth to any other query. It is a finding about reach.
Neither prescription fails at what it specifies: each is stated over the
population and makes no reference to any decision, so neither says which rows a
decision will read. The rows each intervention repairs and the 32 rows the two
decisions actually read do not intersect, which Figure~\ref{fig:intersection}
draws over the population. Population coverage is a property of
how well the data is described. Whether a decision is correct is a property of
its state. The first can improve by any margin without the second moving at all.

\subsection{Contributions}
\label{sec:introduction:contributions}

Neither prescription is claimed as new. Both are long-established practice in
fields that do not cite one another, and no prior work measuring whether either
of them changes a verdict was found. Every contribution below is therefore a
measurement of how far they reach, not a proposal for what to do.

\begin{enumerate}
  \item \textbf{Three measured interventions} against the deployment that
    motivated the prescriptions, each with its method, its figures, and a
    reproduction path from the frozen snapshot alone.
  \item \textbf{A measured production-path decomposition} of the population:
    four paths, distinguished mechanically in code, accounting for all
    $194{,}620$ rows with nothing unclassified --- and separating on an axis the
    prescriptions do not use.
  \item \textbf{A coverage ceiling that is not a vocabulary property.} Three of
    the four candidate vocabularies partition the population identically, and all
    stop at $98.4\%$ --- the figure is $98.4226\%$ in Table~\ref{tab:i1} --- with
    the residual set by an absent record rather than by the choice of terms.
  \item \textbf{An enumeration of the conditions} under which the two decisions
    would have admissible input at all, measured rather than proposed, asserted or
    assumed --- including the observation that the two are blocked for different
    reasons, so a single prescription does not address both.
\end{enumerate}

\subsection{Roadmap}
\label{sec:introduction:roadmap}

Section~\ref{sec:setting} describes the snapshot and the measurement discipline.
Section~\ref{sec:prescriptions} states the two prescriptions as they were written.
Section~\ref{sec:interventions} reports the three interventions.
Section~\ref{sec:result} draws out what they share.
Section~\ref{sec:whynot} enumerates the conditions that would have to change.
Section~\ref{sec:related} situates the work. Section~\ref{sec:limitations} states
what these measurements cannot support, and Section~\ref{sec:conclusion} concludes.

\section{Setting and Method}
\label{sec:setting}

\subsection{What is measured against}
\label{sec:setting:snapshot}

All figures come from a frozen snapshot of the deployment, restored into two
disposable containers. The production deployment was never started. Only read
queries were issued against the restored state, except where an intervention is
itself a program, in which case recorded writes were replayed \emph{through} it
rather than executed against anything.

Before each intervention the restored state was checked against the snapshot
verification script and had to report 33 of 33 passing. A run that did not was
halted rather than adjusted.

Two properties of the arrangement constrain everything that follows.

\paragraph{Grades are not a column.} They were assigned after the fact by an
expression over stored fields, and never written into the snapshot. Adding a
column would mean that later reproductions target a modified snapshot rather than
the recorded one, and the recorded one is the object of study.

\paragraph{No condition was changed to obtain a result.} Where an intervention did
not move a decision, that is reported as the outcome rather than retried under
different settings.

\subsection{The population and how it was produced}
\label{sec:setting:paths}

The two tables the decisions read hold \textbf{194,620} rows. Every row was traced
to the code that produced it. Four production paths account for the population:

\begin{table}[htbp]
\centering
\small
\begin{tabular}{@{}lrp{349pt}@{}}
\toprule
Path & Rows & How a value is produced \\
\midrule
\texttt{stateful} & 69,924 & depends on the previous value: a first-order lag toward a noisy target \\
\texttt{interp} & 330 & a smoothstep kernel between two endpoints, or linear interpolation between anchors \\
\texttt{draw} & 121,296 & a memoryless draw around a reference, scaled by functions of the timestamp \\
\texttt{unattrib} & 3,070 & committed data files whose producing code is not in the repository \\
\midrule
\textbf{Total} & \textbf{194,620} & equals the population; nothing unclassified \\
\bottomrule
\end{tabular}
\caption{Production paths, distinguished mechanically in code.}
\label{tab:paths}
\end{table}

The axis that separates these is not the one the prescriptions assume. What divides
them in code is \textbf{whether a value depends on the previous value}, and among
those that do not, \textbf{whether it interpolates between given points or draws
around a reference}. Whether a value was \emph{observed} divides nothing here: it is
constant across the population, because no observing producer exists in the
repository.

That is worth stating early, because the prescriptions are built on the
observed-versus-derived distinction, and in this deployment that distinction has
only one side.

\subsection{Anonymization}
\label{sec:setting:anonymization}

The deployment is not identified. Table, schema, column, module, function and script
names are replaced by role names; the two verification routines under test are
labelled \textbf{V1} (time-series branch) and \textbf{V2} (relational branch). All
domain-specific vocabulary is removed, as are physical units and values, container
names, ports, paths and commit identifiers. No argument below depends on any of
them.

\textbf{Every measured value is unchanged.} Where a number appears in this paper it
is the number that was measured, unrounded.

\section{The Prescriptions Under Test}
\label{sec:prescriptions}

The companion paper ends with two structural prescriptions. They are stated here in
the form they were given, because the point of this paper is to measure them as
written rather than as improved.

\paragraph{A provenance grade is a structural defense against the error class.}
Marking each row as raw, derived, or interpolated writes the promotion step into
the data itself, so a verification routine cannot mistake the system's own output
for an observation.

\paragraph{A write gateway is where that grade becomes enforceable.} A grade with
independent write paths around it is a convention, and conventions promote
silently. Twelve such paths were recorded in the deployment, none passing through a
validating entry point.

\subsection{What a measurement of these has to answer}
\label{sec:prescriptions:questions}

Both prescriptions are stated over the population --- every row carries a grade,
every write passes an entry point. Neither is stated over the decisions. The
question this paper puts to them is therefore not whether they do what they say,
but whether doing it reaches the place the diagnosis was about.

Three interventions follow from that. Section~\ref{sec:interventions:i3} applies
the grade at the point of consumption, by filtering the verification queries.
Section~\ref{sec:interventions:i1} asks what vocabulary the population actually
requires, rather than assuming the three terms above. Section~\ref{sec:interventions:i2}
implements the single ingress and replays the recorded writes through it.

\subsection{The defect they target is present}
\label{sec:prescriptions:defect}

Nothing here is a test against a system that does not have the problem. In the
snapshot, one verification decision rests on \textbf{24} values of which
\textbf{24} were written by the system itself --- 21 interpolated and 3 stateful,
and \textbf{0} observed. Its relational counterpart returns \textbf{8} rows, all
stateful, and reads \textbf{0 of the 45} rows the writer had just produced for the
same remediation, reporting success on the 8 instead.

\section{Three Interventions}
\label{sec:interventions}

\subsection{I3 --- filtering the verification queries by grade}
\label{sec:interventions:i3}

\paragraph{Method.} V1 and V2 were duplicated with a grade predicate added and
nothing else changed: same window, same thresholds, same row limit. The duplicates
consume only rows graded as observation. The originals were left untouched.

The decision was given three values --- pass, violate, and undetermined --- where
undetermined means the population the decision requires was not met. The original
routines emit only two: their code returns a boolean, so \emph{violate} and
\emph{undetermined} leave by the same exit.

\paragraph{Observed.}

\begin{table}[htbp]
\centering
\small
\begin{tabular}{@{}lll@{}}
\toprule
 & V1 & V2 \\
\midrule
Original result & \textbf{pass} & \textbf{pass} \\
Rows consumed & \textbf{24} of 52 underlying & \textbf{8} \\
\quad of those, observation-graded & \textbf{0} & \textbf{0} \\
\quad composition & 21 interpolated, 3 stateful & 8 stateful \\
Filtered result & \textbf{undetermined} & \textbf{undetermined} \\
\bottomrule
\end{tabular}
\caption{I3. Both decisions move from pass to undetermined.}
\label{tab:i3}
\end{table}

V2 additionally reads \textbf{0 of the 45} rows the writer had just produced for
the same remediation. All 8 rows it returns are stateful output stamped roughly 14
months after the measurement instant, and it reports success on them.

\subsection{I1 --- deciding the grade vocabulary by measurement}
\label{sec:interventions:i1}

\paragraph{Method.} Rather than adopting the three terms the prescription names,
four candidate vocabularies were scored against the population: the three-value
scheme as prescribed; a four-value scheme adding a term for memoryless draws; a
formulation as three boolean axes; and a two-value attributable/unattributable cut.

\paragraph{Observed.}

\begin{table}[htbp]
\centering
\small
\begin{tabular}{@{}lrrrr@{}}
\toprule
Scheme & Classified & Share & Unclassified & Share \\
\midrule
A --- three values & 70,254 & \textbf{36.0980\%} & \textbf{124,366} & \textbf{63.9020\%} \\
B --- four values & 191,550 & \textbf{98.4226\%} & 3,070 & \textbf{1.5774\%} \\
C --- three boolean axes & 191,550 & 98.4226\% & 3,070 & 1.5774\% \\
D --- two values & 191,550 & 98.4226\% & 3,070 & 1.5774\% \\
\bottomrule
\end{tabular}
\caption{I1. Coverage of four candidate vocabularies.}
\label{tab:i1}
\end{table}

Three facts about that table matter more than the coverage figures themselves.

\textbf{B, C and D partition the population identically.} They differ in
expression, not in result; choosing among them is not a choice about coverage.

\textbf{The ceiling is $98.4\%$, and no vocabulary reaches past it.} The
residual $1.6\%$ --- $1.5774\%$ in the table above --- is the unattributable
set. Those rows are unclassifiable in every scheme because the fact needed to
grade them --- what produced the row --- exists neither in the data nor in the
repository. The ceiling is set by an absent
record, not by vocabulary design.

\textbf{Grades are decidable at write time in all four schemes}, and only partly
decidable afterwards from stored fields, because the post-hoc expression needs two
window constants that appear nowhere in the data.

\subsection{I2 --- a single write ingress that requires a grade}
\label{sec:interventions:i2}

\paragraph{Method.} An ingress was implemented as a validator, and the recorded
writes were replayed through it row by row. The production system was not modified.
The contract has four stages, checked in order: \textbf{S1} is the producer of this
row registered; \textbf{S2} does that producer declare a production path;
\textbf{S3} does the path map to a value in the vocabulary; \textbf{S4} are the
required fields present. Only producers confirmed in code were registered, so the
committed data files could not be --- there is no producing code to name.

\paragraph{Observed --- full replay.}

\begin{table}[htbp]
\centering
\small
\begin{tabular}{@{}lr@{}}
\toprule
 & Rows \\
\midrule
Replayed & \textbf{194,620} \\
Admitted & \textbf{191,550} \\
Rejected & \textbf{3,070} \\
\quad at S1 & \textbf{3,070} \\
\quad at S2, S3, S4 & \textbf{0, 0, 0} \\
\bottomrule
\end{tabular}
\caption{I2. Every rejection occurs at the first of four stages.}
\label{tab:i2}
\end{table}

Admitted rows by grade: 69,924 stateful, 121,296 draw, 330 interpolated, and
\textbf{0 observation}.

One thing follows from the stage column, and one thing does not. Every
rejection occurs at \textbf{S1}: no row is refused for a reason other
than an unnameable producer. Filling in data would not admit them; only
finding the producer would. What does not follow is anything about
\textbf{S2}, \textbf{S3} and \textbf{S4}. The stages are checked in
order, so no rejected row reached them; the $191{,}550$ admitted rows
passed all four. A stage that refuses nothing has not thereby been
shown to have inspected anything. A zero was recorded at each of the
three and read as a property of those stages. It is not.

\paragraph{Observed --- the rows the decisions consume.} The 32 rows V1 and V2 read
were replayed through the same ingress: \textbf{32 replayed, 32 admitted, 0
rejected} --- 21 interpolated, 11 stateful, and \textbf{0} observation-graded.

The intersection of the rows the ingress rejects and the rows the decisions read is
empty. Both decisions remain undetermined.

\section{The Common Result}
\label{sec:result}

All three interventions leave both decisions without admissible input.

\begin{table}[htbp]
\centering
\small
\begin{tabular}{@{}lp{96pt}ll@{}}
\toprule
Intervention & What it changed & V1 & V2 \\
\midrule
I3 --- grade filter & the admissible set & pass $\rightarrow$ \textbf{undetermined} & pass $\rightarrow$ \textbf{undetermined} \\
I1 --- wider vocabulary & classified $36.0980\% \rightarrow 98.4226\%$ & \textbf{unchanged} & \textbf{unchanged} \\
I2 --- single ingress & refused 3,070 writes & \textbf{unchanged} & \textbf{unchanged} \\
\bottomrule
\end{tabular}
\caption{Three interventions, two decisions.}
\label{tab:common}
\end{table}

\begin{figure}[htbp]
\centering
\begin{tikzpicture}[x=1mm, y=1mm, font=\small]

  \def\Wdraw{81}   
  \def\Wunat{2}    

  \fill[black!12] (0,0) rectangle (\Wdraw,22);
  \fill[black!55] (\Wdraw,0) rectangle (83,22);
  \draw[semithick] (0,0) rectangle (130,22);
  \draw[semithick] (\Wdraw,0) -- (\Wdraw,22);
  \draw[semithick] (83,0) -- (83,22);

  \node[anchor=south west, font=\small\itshape] at (0,29)
    {frozen snapshot --- 194{,}620 rows};

  \node[align=center, text width=70mm] at (40.5,11)
    {\textbf{draw} --- 121{,}296 rows\\[1pt] \footnotesize I1 names these};
  \node[align=center, text width=40mm] at (106.5,16)
    {\textbf{already named}\\ --- 70{,}254 rows};

  \draw[semithick] (82,22) -- (82,27);
  \node[anchor=south, font=\footnotesize, align=center] at (82,27)
    {\textbf{unattributable} --- 3{,}070\\ I2 refuses these};

  \draw[very thick] (101,2.5) rectangle (109,8.5);
  \node[font=\footnotesize] at (105,5.5) {\textbf{32}};
  \draw[semithick] (105,2.5) -- (105,-7);
  \node[anchor=north, align=center, font=\footnotesize] at (105,-7)
    {the 32 rows V1 and V2 read --- 24 and 8,\\
     none of them observation-graded\\
     \emph{(box enlarged; the bands are to scale)}};

  \draw[semithick] (0,-2) -- (0,-4) -- (83,-4) -- (83,-2);
  \node[anchor=north, font=\footnotesize] at (41.5,-4)
    {unclassified under the prescribed vocabulary --- 124{,}366 rows};

\end{tikzpicture}
\caption{The snapshot divided by what each intervention reaches: I1 names the
$121{,}296$ \texttt{draw} rows, I2 refuses the $3{,}070$ unattributable rows, and
the 32 rows the two decisions read lie inside neither. Between them the two
interventions move every row the prescribed vocabulary could not name ---
$124{,}366$ of them --- and not one of the 32, which is why an improvement in
coverage of this size leaves both decisions exactly where they were.}
\label{fig:intersection}
\end{figure}
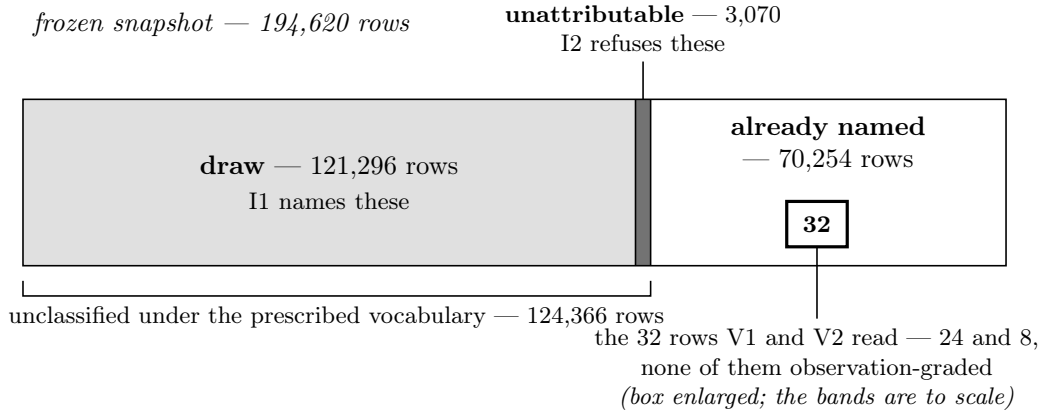

I1 absorbed $121{,}296$ rows that the prescribed vocabulary could not name. I2
refused $3{,}070$ writes that could not be attributed to any producer. Neither
touched the 32 rows on which the two decisions actually rest: those 32 contain no
\texttt{draw} rows and no unattributable rows. Figure~\ref{fig:intersection}
draws the three sets over the snapshot.

%

\medskip
\noindent\textbf{What each intervention repaired and what the decisions read do not
overlap.} Population coverage is a property of the description of the data. The
correctness of a decision is a property of its state. Improving the first does not
move the second, and the size of the improvement gives no indication of whether it
will.
\medskip

I3 is the apparent exception, and it is worth being exact about why it is not one.
It does change both decisions --- from pass to undetermined --- but not by supplying
better input. It changes them by removing all input: the admissible set it enforces
is empty in this deployment, so the decisions stop rather than improve. The change
is real and is arguably the correct behaviour, since a decision with no admissible
evidence should not report success. But it is not the prescription reaching the
decision with something to decide on.

The pattern is the same one the companion paper catalogues, turned on the
prescriptions rather than on the system. A grade covering $98.4\%$ of rows
(Table~\ref{tab:i1}) is a description of the data. An ingress that refuses every unattributable write is a
description of the write path. Neither is, by itself, evidence about the state a
verification routine will read --- and in this deployment neither turned out to be
evidence about it at all.

\section{Why the Decisions Do Not Move}
\label{sec:whynot}

If none of the three interventions gives the decisions admissible input, the useful question is what
would. The conditions below are enumerated from measurement. They are necessary
conditions observed to be unmet, not a proposed design, and they are not ranked.

\paragraph{K1 --- the admissible set must contain a grade that occurs in the query
window.} Observation-graded rows number \textbf{0} across the whole population, so
an admissible set of \texttt{\{observation\}} can never be met, under any vocabulary
and behind any ingress. This is the condition all three interventions leave
untouched.

\paragraph{K2 --- V1 needs only K1.} Its window holds \textbf{45} interpolated
buckets against a threshold of \textbf{5}, plus one stateful bucket.

\begin{table}[htbp]
\centering
\small
\begin{tabular}{@{}lrl@{}}
\toprule
Admissible set & Buckets returned & Result \\
\midrule
observation & 0 & undetermined \\
stateful & 1 & undetermined \\
interpolated & \textbf{45} & has input \\
interpolated + stateful & 46 & has input \\
\bottomrule
\end{tabular}
\caption{K2. Admit the interpolated grade and V1 has input.}
\label{tab:k2}
\end{table}

\paragraph{K3 --- V2 needs more than K1.} Its query orders by time descending and
takes the first 8 rows. The first row under verification sits at \textbf{rank
69,789}; the query reaches \textbf{rank 8}. The gap is \textbf{69,781} rows.
Widening the admissible set alone returns nothing, because $69{,}788$ later-stamped
rows stand in front of the 45 rows being verified.

\paragraph{K4 --- neither query bounds its window above.} In V1's window,
\textbf{10} rows fall outside the remediation window and enter the decision anyway;
3 of the 24 rows the decision consumed came from there.

\paragraph{K5 --- the 3,070 unattributable rows pass under no condition.} No
admissible set, vocabulary or ingress admits them. The only path is to find what
produced them, and that fact is absent from both the data and the repository.

\subsection{The two decisions are blocked differently}
\label{sec:whynot:asymmetry}

K2 and K3 are the pair worth separating. V1 is short of an admissible grade and
nothing else: satisfy K1 in a way that admits interpolated rows and it has input.
V2 is short of an admissible grade \emph{and} unable to reach the rows in question
even if it had one, because the ordering delivers other rows first.

So the two decisions, which sit in the same routine and were treated as one case by
the diagnosis, do not respond to the same change. A prescription stated over the
population does not distinguish between them, and the measurement does.

This also explains why the intervention that changed the most rows changed the
least here. I1 moved $121{,}296$ rows from unnameable to named, and every one of
them lies outside both query windows. Coverage and reach are independent, and in
this deployment they turned out to be uncorrelated.

\section{Related Work}
\label{sec:related}

Neither prescription measured here is new. Both have been standard practice for
years, in fields that do not cite each other, and saying so first is necessary:
what is offered is a measurement, and a measurement of a novel prescription
would be worth less.

\paragraph{Grading a value by how it was obtained.}
Marking each record with the kind of evidence behind it is an established
convention with two decades of deployment. The Evidence and Conclusion Ontology
provides over fifteen hundred terms describing types of evidence and assertion
methods, structured on two roots --- the evidence itself, and whether the
assertion was made by a human or produced
automatically~\cite{giglio2019eco}. In official statistics, the SDMX observation
status code list attaches a status to each observation: normal, estimated,
imputed, provisional, forecast, and a family of missing-value
codes~\cite{sdmx2019obsstatus}. And provenance-aware storage grades its own
metadata three ways: a PASS ``distinguishes between internally-collected
provenance, annotations, and application-generated provenance so queries can
specify which attribute types to consider''~\cite{muniswamy2006pass}. The four
grades of Section~\ref{sec:prescriptions} are a rediscovery of this family, not
an addition to it.

\paragraph{What an absent grade is taken to mean.}
One detail of the statistical standard is worth stating on its own, because it
is this deployment's failure written into an international convention. The code
list provides no value for an observation whose origin is unknown, and specifies
that where no status has been associated with an observation, the code for a
normal value may be assumed --- a code whose own definition is that the source
agency assigns sufficient confidence to the
observation~\cite{sdmx2019obsstatus}. The absence of a record about a value is
thereby promoted to a claim that the value is ordinary and trusted. The $3{,}070$
rows of Section~\ref{sec:interventions} that no vocabulary can classify would,
under that convention, be classified as normal.

\paragraph{Capture at one point.}
The second prescription --- route every write through one place --- is the
founding argument of provenance-aware storage. Muniswamy-Reddy et al.\ argue
that provenance belongs in the storage system rather than in applications,
because collection is then transparent and produces ``a level of meta-data
completeness difficult or impossible to achieve with application-level
solutions''~\cite{muniswamy2006pass}. Pasquier et al.\ move the same argument
into the kernel, capturing at Linux Security Module hooks on the path between
any kernel object and a process~\cite{pasquier2017camflow}. That architecture is
prior art and is not claimed here.

\paragraph{What neither capture system does is refuse.}
Both observe; neither admits or rejects. PASS intercepts system calls and
translates them into provenance records, and its seventeen-step account of
collecting provenance for a single command contains no step that can fail the
call~\cite{muniswamy2006pass}. CamFlow is more pointed, because it sits in the
one Linux subsystem whose hooks genuinely can fail a system call, and then
arranges not to: its hook is called last, the authors write, to avoid recording
flows that another module subsequently blocks~\cite{pasquier2017camflow}. Its
policy engine discards records, not operations. And where PASS meets data whose
origin it cannot see --- which it calls opaque provenance, arising when data
comes from a user, another machine, or a file system that is not
provenance-aware --- its three remedies are best-effort deduction, voluntary user
annotation, and application-supplied provenance~\cite{muniswamy2006pass}. A row
with no attributable origin is admitted by both systems. The ingress of
Section~\ref{sec:interventions} refuses it, and that difference is what the
$3{,}070$ refusals measure.

There is an inversion here worth stating plainly. PASS observed in 2006 that it
is not possible to automatically collect provenance the system never
sees~\cite{muniswamy2006pass}, and chose to admit such data and invite an
annotation. The measurement reported here is what the other branch buys. All
$3{,}070$ refusals fall at the first stage, producer registration; the three
later stages never fire. A chokepoint cannot manufacture attribution for data
that arrived without it. It can only decline.

\paragraph{What is recorded is derivation, not the kind of origin.}
Both systems record a causal graph. PASS defines provenance as a description of
the execution history that produced an object, with complete provenance being
the transitive closure over references, captured as a
DAG~\cite{muniswamy2006pass}. CamFlow extends the W3C provenance data model,
with entities, activities and agents joined by used, generated-by and
associated-with edges~\cite{pasquier2017camflow}. Both answer \emph{what did
this come from}. Neither carries a value stating \emph{what kind of thing this
is} --- observed, derived, interpolated, unattributed --- travelling with the
datum and checkable at admission. The two axes are complementary and the second
is the one Section~\ref{sec:prescriptions} tests.

\paragraph{No measurement found of whether a grade reaches a decision.}
This is the gap the paper occupies, and three of the works above say so in their
own words. ECO states that its terms do \emph{not} indicate any level of quality
or confidence in the evidence or the assertion, only the type of evidence, and
that a parallel system for capturing confidence is therefore
needed~\cite{giglio2019eco}. The SDMX list ties no code to any gate, threshold
or refusal; its single instruction to a consumer is that a user should be aware
of low reliability~\cite{sdmx2019obsstatus}. CamFlow's evaluation is scoped in
three questions --- maintainability, application performance, and the volume of
data generated --- and where the completeness of what is captured comes up, the
authors write that to their knowledge no study of such a nature has been
completed~\cite{pasquier2017camflow}. PASS reports space overhead, time overhead
and query latency, and no other evaluation dimension was found in
it~\cite{muniswamy2006pass}. To the best of the author's knowledge, every
evaluation in this literature measures the cost of the discipline and assumes
its utility. This paper measures the utility.

\paragraph{Coverage as a proxy.}
The result reported here is of a kind the testing literature has established in
another setting. Inozemtseva and Holmes measured $31{,}000$ suites across five
Java systems of up to $724{,}000$ lines and found the correlation between
coverage and fault detection low to moderate once the number of test cases is
controlled for, concluding that coverage should not be used as a quality
target~\cite{inozemtseva2014coverage}. Section~\ref{sec:interventions} measures
the analogous quantity over a data population: classified coverage rises from
$36.1\%$ to $98.4\%$ and neither decision moves. The difference is one of
degree in the strongest sense. A low-to-moderate correlation still contains
cases in which improving the proxy improves the outcome; what
Section~\ref{sec:result} reports is an intersection of size zero between the
rows an intervention repairs and the $32$ rows the decisions read.

\paragraph{Executed, and actually consulted.}
Schuler and Zeller separate the statements a test executes from those whose
results reach an oracle; their opening example attains $83\%$ statement coverage
and $0\%$ checked coverage because no computed result flows into a
check~\cite{schuler2011checked}. The relation between $194{,}620$ rows and the
$32$ a decision consumes is the same relation. What differs is the instrument:
checked coverage is computed by a dynamic slicer in a single instrumented run,
whereas the figure here was obtained by reading two verification queries and
what they executed, once each, by hand.

\paragraph{Passing without deciding anything.}
Beer et al.\ formalize vacuity in temporal model checking: a formula may be
valid because the antecedent of its implication is not satisfiable in the model,
so that a valid formula can hide a real problem rather than attest to its
absence. They report from practice at IBM that in first formal verification runs
of a new hardware design typically $20\%$ of formulas are found trivially valid,
and that trivial validity always indicated a real problem in the design, its
specification, or its environment~\cite{beer2001vacuity}. Intervention I3
produces a result of this shape. Filtering the verification queries to
observation-graded rows returns undetermined for both decisions, but no row in
the population carries that grade, so the filter's admitting branch never
executed. The outcome is a property of an empty set rather than a discrimination
the filter performed, and Section~\ref{sec:whynot} treats it as such.

\paragraph{Validating what enters a pipeline.}
Breck et al.\ describe a data validation system deployed at Google as part of
TFX, validating several petabytes of production data per day against a
generalized schema~\cite{breck2019datavalidation}. The gap between it and the
prescriptions under test is that paper's own premise: the pipeline receives data
in a raw-value format that strips out the semantic information which would
identify an error, so a value can be valid for its type and carry nothing about
how it arose. Their answer constrains the value; the prescriptions here
constrain the origin. Neither substitutes for the other, and a schema would have
accepted every row the ingress refused.

\paragraph{What remains particular.}
The prescriptions are old and the architecture is older. Three things are
particular to what is reported. The \emph{refusal}: both capture systems admit a
record whose origin they cannot establish, and the measurement here is of the
branch they declined to take. The \emph{null result}: nothing found in the
literature measures whether a provenance grade changes a decision, and three of
the works above state as much themselves. And the \emph{shape of the null}: not
a weak correlation but an empty intersection, over a population accounted for in
full, with the two decisions blocked for different reasons so that no single
prescription among those tested reaches both.

\section{Limitations}
\label{sec:limitations}

These bound what the paper may conclude. Several of them weaken its claims
materially and are stated anyway.

\begin{enumerate}

\item \textbf{One deployment, two decisions. This is not a sample.} The frozen
snapshot supports exactly one decision per verification routine, so every result
above rests on two observations. The count is a property of the design rather than
of the measurement effort --- the remediation window is pinned to fixed constants,
the writer rewrites that same window on every call, and its generator is re-seeded
identically each time. Recomputing the curve from the constants alone reproduces
the stored rows \textbf{270 of 270 exactly}, so repeated invocations are replays of
one computation, not independent observations. No amount of running produces a
third.

\item \textbf{Observation-graded rows are 0, so no filter ever executed its
admitting path.} Every filtered decision in this paper reached undetermined by way
of an empty admissible set. \emph{Nothing here is evidence about how a grade filter
behaves when observation-graded rows exist} --- whether it admits the right ones,
rejects the right ones, or decides correctly on them. That path was never
exercised, and the interventions cannot speak to it.

\item \textbf{The ingress is a replay validator, not a deployed gateway.} The
production system was not modified. Recorded writes were streamed through a program
implementing the contract. Nothing here bears on its behaviour in operation: batch
semantics, performance, transaction boundaries, failure modes under concurrent
writers, and the cost of enforcement were all untouched.

\item \textbf{The remediation routine was not executed.} The determinism check
answers the repeat-invocation question without running it, but that leaves one
thing unverified in this paper and in the measurements that preceded it: whether
the write path completes end to end. Neither this work nor the earlier work checked
it.

\item \textbf{The author measured their own prescriptions.} The diagnosis, the
prescriptions, the interventions and the scoring of those interventions are one
body of work, carried out by one person over a single period, with no second party
at any stage. The result is unflattering to the prescriptions, which is some
protection against motivated scoring but not much, and none at all against a blind
spot in what was chosen to measure.

\item \textbf{A null result about reach is not a result about value.} What was
measured is that three interventions do not give these two decisions admissible input in this
deployment. Whether the $3{,}070$ refused writes or the $121{,}296$ newly nameable
rows matter to other queries, other decisions, or later work was not examined, and
this paper should not be read as evidence that they do not.

\item \textbf{The enumerated conditions are necessary, not sufficient.} K1 through
K5 are conditions observed to be unmet. That satisfying K1 and K2 gives V1 input is
measured; that it would then decide \emph{correctly} is not, and no measurement
here supports it.

\item \textbf{The unattributable rows remain unattributed.} $3{,}070$ rows have no
identifiable producer, and this work did not find one. Their production method is
as unknown at the end as at the start.

\item \textbf{Neither prescription is this work's invention, the practice they come
from has now been read, and one adjacent line has not.} Marking each record with
the kind of evidence or the origin standing behind it is an established practice
with its own vocabularies and standards in other fields, and so is funnelling every
write through one enforced point of capture rather than trusting each writer to
comply. That practice has now been read in full, and Section~\ref{sec:related}
positions the prescriptions against it --- the grading standards, the two capture
systems, and, in three cases, those works' own statements about what they leave
unmeasured. This paper neither proposes the two prescriptions nor claims them as
new; it measures how far they reach in one deployment, which is a question about
reach and not about novelty. What that reading changed is the framing and not the
figures: no measurement in this paper was revised in consequence. One line
identified in the same search remains unread --- the safety-engineering literature
on the distance between work-as-imagined and work-as-done, where the gap between a
written procedure and the executed one has its own decades of treatment under its
own vocabulary. Whether the reach result reported here is already stated there is
unknown, and a reader who knows that literature should read
Section~\ref{sec:related} as partial in exactly that place.

\item \textbf{The deployment cannot be inspected by the reader.} It is anonymized,
and the figures are checkable only against a snapshot the author produced and
verified. Internal consistency and reproduction from that snapshot are available; independent
confirmation that the snapshot represents the deployment is not.

\end{enumerate}

\section{Conclusion}
\label{sec:conclusion}

Two prescriptions were measured against the deployment that motivated them, over a
frozen snapshot of $194{,}620$ rows and the two verification decisions it supports.

Filtering the verification queries by grade turns both decisions from pass to
undetermined --- by removing all their input rather than by improving it, since the
admissible set it enforces is empty here. Widening the grade vocabulary raises
classified coverage from $36.1\%$ to $98.4\%$ (Table~\ref{tab:i1} carries both
to four decimal places), and three of the four candidate vocabularies turn out to partition the population identically, all
stopping at the same ceiling, which is fixed by an absent record rather than by the
choice of terms. A single ingress requiring a grade refuses $3{,}070$ of the recorded
writes, every one of them at the first of its four stages. The three
later stages recorded zero refusals, which is not a statement about
them.

Neither of the last two changes either decision. The 32 rows the decisions read
were all admitted by the ingress, contain no rows the wider vocabulary newly names,
and include no observation-graded row. What each intervention repaired and what the
decisions read do not intersect.

The conditions under which the decisions would have admissible input were
enumerated from measurement rather than proposed, and they are not the same for
both: one decision is short of an admissible grade, while the other is short of an
admissible grade and also unable to reach the rows it is verifying, by a margin of
$69{,}781$ ranks. A prescription written over the population does not distinguish
between those two situations.

None of this shows that provenance grades or write gateways are without value, and
Section~\ref{sec:limitations} is explicit about how little two decisions in one
deployment can carry. What it does show is that coverage of a population and reach
to a decision are separate quantities, that the first can improve by a large margin
while the second does not move, and that the distance between them is measurable.

\bibliography{refs}

\end{document}